# Actionable forecasting as a determinant of function in noisy biological systems

Jose M. G. Vilar[1,2,*] and Leonor Saiz[3,*]

[1] Biofisika Institute (CSIC, UPV/EHU), University of the Basque Country (UPV/EHU), P.O. Box 644, 48080 Bilbao, Spain

[2] IKERBASQUE, Basque Foundation for Science, 48011 Bilbao, Spain

[3] Department of Biomedical Engineering, University of California, 451 E. Health Sciences Drive, Davis, CA 95616, USA

[*] Correspondence: lsaiz@ucdavis.edu; j.vilar@ikerbasque.org

## Abstract

Continuous adaptation to variable environments is crucial for the survival of living organisms. Here, we analyze how adaptation, forecasting, and resource mobilization towards a target state, termed actionability, interact to determine biological function. We develop a general theory and show that it is possible for organisms to continuously track their optimal state in a dynamic environment by adapting towards an actionable target that incorporates just current information on the optimal state and its rate of change. If the environmental information is precise and readily actionable, it is possible to implement perfect tracking without anticipatory mechanisms, irrespective of the adaptation rate. In contrast, predictive functions, like those of circadian rhythms, are beneficial if sensing the environment is slow or unreliable, as they allow better adaptation with fewer resources. To explore potential actionable forecasting mechanisms, we develop a general approach that implements the adaptation dynamics with forecasting through a dynamics-informed neural network.

# Introduction

In the search for general biological principles, there has been extensive research on network structure, decision making, and evolution, among others [1-6]. The key emergent results are centered on the idea that biological systems should function reliably, and often optimally, in their environment. The role of forecasting capabilities of biological systems has not been so extensively studied. At the computational level, predicting the evolution of complex systems from continuously sensed real-time data to act upon them and control their behavior is a central challenge across multiple significant problems in biology, society, the environment, industry, human health, and many other domains [7-12]. A prominent example of inherently predictive behavior in biological systems is provided by circadian rhythms [13]. They are internal processes that regulate various physiological and behavioral functions in organisms to parallel the recurring 24-hour daily cycle. In most cases, the recurring patterns persist even when the organisms are placed under controlled constant conditions. The generation of recurrent rhythmic patterns that mimic the expected environment has usually been thought of as an anticipatory mechanism that helps adaptation to daily environmental changes [14, 15]. However, the benefits that the inherent predictive behavior can provide to biological systems and their interplay with actionability, namely, the mobilization of resources based on the anticipated environmental state, have remained largely unexplored.

Here, we develop a general theory to capture the effects of forecasting and actionability in the context of continuous adaptation. Unraveling this connection is important because biological systems naïvely adapt toward a potentially optimal state with a delay [16, 17]. The longer the delay, the farther the system is from the optimal state. Fast adaptation, however, requires higher reaction rates and mobilization of resources, which is generally costly and detrimental [18, 19]. Therefore, actionability requires a tradeoff between speed and resources. Here, the focus is on how organisms can modify this tradeoff towards better adaptation with fewer resources.

Our results show that adapting towards a function of the current environment and its rate of change, which we termed actionable target, rather than to the optimal state itself allows the system to precisely track a changing environment without delay. If the information about the current environment is precise and readily actionable, anticipatory mechanisms are not needed to implement perfect tracking, irrespective of the adaptation rate. Predictability capabilities of circadian rhythms, however, could be advantageous with unreliable and slow sensing of the environment. Explicitly, to study these tradeoffs, we focus on daily and seasonal clocks as well as on the current and recent environment as the sources of information available for the adaptation processes. The main difference between these two types of sources is the instantaneous endogenous information provided by the clocks. In contrast, exogenous environmental information needs to be sensed and relayed through diverse signaling pathways until it is ready to be actionable, which is not instantaneous. As an explicit quantitative environment, we consider solar radiation on the Earth's surface. This quantity is directly relevant to the metabolism of photosynthetic organisms such as plants and cyanobacteria as well as to the behavior of virtually any system that is coupled to the environment, such as organisms and communities that follow the day-night cycle [20].

To implement the adaptation dynamics together with the forecasting approach, we developed a dynamics-informed neural network (DINN) [21]. This approach allows us to test the performance of different actionable forecasting strategies coupled with the dynamics of adaptation.

## Results

### Theory

At the cellular level, the state of the system is in general determined by multiple variables, such as ATP levels, number of proteins, etc. We focus on the quantification of one of these variables. Explicitly, we consider the cellular state at time $t$ described by $y_t$. For a given environment at time $t$, there is an optimal cellular state $y_t^o$ that maximizes the growth rate. Near the optimal state, the growth rate is given by $r_t = r_t^o - c_t (y_t - y_t^o)^2$, with $c_t \equiv -d^2 r_t/(dy_t)^2$, which is positive, and $r_t^o$ being the maximum growth rate. This dependence is important to define the optimization problem. Therefore, the objective is to minimize the mean square error (MSE) between the optimal and actual state over time, which is mathematically defined as $e_{MSE} = \lim_{T \to \infty} \frac{1}{T} \int_0^T (y_t - y_t^o)^2 dt$. It is also useful to consider the root mean square error (RMSE) defined as $e_{RMSE} = \sqrt{e_{MSE}}$, which has the same units as the quantification of the cellular state.

We consider the dynamics given by

$$\frac{dy_t}{dt} = b_t(f_t - y_t),$$

(1)

where $b_t$ is the adaptation rate of $y_t$ towards a function $f_t$ that depends on the environment. This equation can straightforwardly be solved as

$$y_t = y_0 e^{-\int_0^t b_s ds} + \int_0^t e^{-\int_s^t b_z dz} b_s f_s ds.$$

(2)

Naïve adaptation would correspond to adaptation of the cellular state to the optimal value at a given time, i.e., $f_t = y_t^o$. In this scenario, $y_t$ will be different from $y_t^o$.

It is useful to write $f_t$ as $f_t - y_t^o + y_t^o$, substitute it in Eq. (2), and integrate by parts the term $e^{-\int_s^t b_z dz} b_s y_s^o$, which makes use of the identity $\frac{d}{ds} e^{-\int_s^t b_z dz} y_s^o = e^{-\int_s^t b_z dz} b_s y_s^o + e^{-\int_s^t b_z dz} \frac{d}{ds} y_s^o$. The result

$$y_t = y_t^o + (y_0 - y_0^o) e^{-\int_0^t b_s ds} + \int_0^t e^{-\int_s^t b_z dz} b_s \left( f_s - y_s^o - \frac{1}{b_s} \frac{d}{ds} y_s^o \right) ds$$

(3)

shows that it is possible to perfectly track a changing optimal state, so that $y_t = y_t^o$ after the initial transient, if the system relaxes towards

$$f_t = y_t^o + \frac{1}{b_t} \frac{d}{dt} y_t^o,$$

(4)

which we have termed the actionable target. Therefore, precise continuous tracking requires information on the optimal state, its changes, and the adaptation rate. The faster the adaptation rate, the smaller the dependence on the derivative.

Precise tracking does not in principle require information about the future or forecasting approaches. However, this property requires no delays in relaying the optimal state information. If there is a delay $\Delta t$ and the approach is applied straightforwardly, namely, $f_t = y_{t-\Delta t}^o + \frac{1}{b_t} \frac{d}{dt} y_{t-\Delta t}^o$, the system would track the delayed optimum value as $y_t = y_{t-\Delta t}^o$. Because $y_t^o - y_{t-\Delta t}^o \simeq \frac{d}{dt} y_t^o \Delta t$, the value of the MSE, $e_{MSE} \simeq \Delta t^2 \left\langle \left( \frac{d}{dt} y_t^o \right)^2 \right\rangle$, scales proportionally to the square of the delay and the average of the square of the rate of change of the optimal state.

In situations with delays and variable environments, relaxing towards an estimate of the current actionable target could potentially be more efficient than relaxing towards the actual delayed value. Accurately predicting the actionable target is crucial because the system would relax towards $f_t = \hat{y}_t^o + \frac{1}{b_t} \frac{d}{dt} \hat{y}_t^o$, where the hat indicates that the value $\hat{y}_t^o$ is a forecast of $y_t^o$ from past values.

**Validation**

To put the theory in context, we consider adaptation to the normalized hourly changes in solar radiation on the Earth's surface. As a normalization factor, we use the maximum radiation. As a representative location, we selected latitude 45° N and longitude 0° E. The specific latitude has marked seasonal effects superimposed on daily changes as well as weather patterns (Fig. 1). The longitude corresponds to the Greenwich meridian, for which the Coordinated Universal Time (UTC) corresponds to the mean solar time. The values were obtained from the PVGIS v5.2 database [22]. It uses satellite data to compute radiation at hourly resolution from the years 2005 to 2020. We validate the approach with the data from the years 2015 to 2020. We keep the data from the years 2005 to 2014 for training.

The approach is embedded into a predictive framework through a dynamics-informed neural network (Figure 2). For the adaptation dynamics, we rely on the discretized integral representation [Eq. (2)] with a time-independent adaptation rate constant ($b_t = b_0$) and without the transient term (long-term behavior), which we implement as a convolutional layer of the neural network. The discretization follows the sampling of the solar radiation. Explicitly, we use $y_t = \sum_{i=0}^{M} w_i f_{t-ih}$ with the kernel $w_i$ given by $w_0 = \frac{1}{2} b_0$ and $w_{i>0} = b_0 e^{-b_0 i h}$. Here, $M$ is the size of the kernel and $h$ represents a 1-hour interval. This result is obtained from the trapezoidal rule for numerical integration after a change of variables in the convolution of Eq. (2). We use this approach with a single-neuron linear layer with fixed weights to implement numerical approaches as well as with a trainable deep neural network (DNN) to capture potentially more complex predictive non-linear relationships.

The naïve adaptation mechanism, with $f_t = y_t^o$, leads to a substantial delay in tracking the optimal state (Fig. 3A), resulting in an RMSE of 0.147. Incorporating the optimal state rate of change through a first-order approximation of the derivate, $f_t = y_t^o + \frac{1}{hb_0}(y_t^o - y_{t-h}^o)$, provides much better tracking than the naïve mechanism (Fig. 3B) with an RMSE of 0.027. Trying to improve the estimation of the derivative through a centered second-order approximation, $f_t = y_t^o + \frac{1}{2hb_0}(y_{t+h}^o - y_{t-h}^o)$, is not possible because it will require values from the future. Relying only on current and past values, a second-order backward difference approximation of the derivative, $f_t = y_t^o + \frac{1}{2hb_0}(3y_t^o - 4y_{t-h}^o + y_{t-2h}^o)$, further improves the tracking (Fig. 3C) with an RMSE of 0.013. This result shows that if the actionable target is not delayed, there is accurate tracking without anticipatory mechanisms, even when the environment is noisy.

**Delayed inference and extrapolation**

The main limitation of this type of tracking lies in the ability to infer the value of the actionable target at time $t$. Explicitly, if only the past is actionable, namely, if there is a delay, the tracking ability is substantially reduced, for the naïve case, $f_t = y^o_{t-h}$, with RMSE of 0.195 (Fig. 4A) and for the second backward difference approximation of the derivative, $f_t = y^o_{t-h} + \frac{1}{2hb_0}(3y^o_{t-h} - 4y^o_{t-2h} + y^o_{t-3h})$, with RMSE of 0.079 (Fig. 4B). Note that in the latter case the RMSE is essentially the square root of the average value of $(y^o_t - y^o_{t-h})^2$, which in this case is 0.082. We also considered the second order extrapolation of the optimal state and its derivative as $\hat{y}^o_t = y^o_{t-h} + \frac{1}{2}(3y^o_{t-h} - 4y^o_{t-2h} + y^o_{t-3h})$ and $\frac{d}{dt}\hat{y}^o_t = \frac{1}{2h}(5y^o_{t-h} - 8y^o_{t-2h} + 3y^o_{t-3h})$, which leads to an RMSE of 0.073 (Fig. 4C). Here, we have used explicitly $f_t = \hat{y}^o_t + \frac{1}{b_t}\frac{d}{dt}\hat{y}^o_t$. In this case, linear extrapolation does not significantly improve the precision of the tracking because of the inherently noisy nature of the environment.

If the system has a recurrent component, such as those of daily cycles, it would be possible to estimate the optimal state and its change considering also values from one day earlier, namely, as $\hat{y}^o_t = y^o_{t-h} + y^o_{t-24h} - y^o_{t-24h-h}$ and $\frac{d}{dt}\hat{y}^o_t = \frac{1}{2h}(y^o_{t-24h+h} - y^o_{t-24h-h})$. The resulting tracking has an RMSE of 0.087 (Fig. 5A). This type of approach, which can be improved to consider multiple days back in time, as in the Holt-Winters' seasonal method [23], provides the system with periodic information. When the results of the delayed and the recurrent approaches are averaged together, the RMSE is reduced to 0.066 (Fig. 5B), which is lower than the values obtained for each of them separately. The tracking error can be reduced further by considering the extrapolated instead of the delayed approach in the average with the recurrent information, which leads to an RMSE of 0.057. Such consistent error reductions upon averaging different estimations indicate that random fluctuations play a fundamental role in preventing the system from tracking the optimal value when forecasting is needed.

Overall, the results show that, indeed, it is possible to substantially improve the naïve mechanism considering adaptation to the delayed actionable target, which can be further improved through forecasting, or anticipatory, mechanisms for the current optimal state and its time-derivative contribution (Figs. 4 and 5). The forecasting mechanisms used so far should be considered as a baseline since they are linear, account for the recent past, and incorporate only the recurrences of 1 day before.

**An integrated actionable forecasting strategy**

Through the previous analyses, we have shown that in general there is short- and long-term information that can contribute to predicting changes in the optimal state. To generally address these dependencies, we consider a general Deep Neural Network (DNN) framework that explicitly incorporates short-term, long-term, and recurrent dependencies. The motivation for using a DNN is their ability to encode general functional dependencies between variables, their trainability properties, and their computational equivalence to biomolecular networks. Indeed, multiple results have shown explicitly that biomolecular networks, including enzymatic, signal transduction, and gene regulatory networks, can perform computations equivalent to those of artificial neural networks and other architectures in machine learning [24-27].

The potential role of recurrent effects enters the approach through a functional dependency on the time of the day (circadian) and the time of the year (circannual). Multiple organisms, from microorganisms as simple as bacteria to humans, have indeed biomolecular mechanisms to tackle daily changes in environmental conditions. Circannual rhythms are needed to regulate physiological and behavioral processes over the changes that organisms experience through seasons, such as temperature and day-length changes [28, 29]. To account for these recurrent events, we consider explicitly a clock that performs a circular motion in the unit circle in phase space as $\mathbf{c}_t^d = \left(\sin\frac{2\pi t}{\tau_D}, \cos\frac{2\pi t}{\tau_D}\right)$ for the daily changes and $\mathbf{c}_t^a = \left(\sin\frac{2\pi t}{\tau_Y}, \cos\frac{2\pi t}{\tau_Y}\right)$ for annual effects. Here, $\tau_D$ is the day length (24 hours) and $\tau_Y$ is the year length (265.25 days). We do not delve into the potential biomolecular mechanisms for the clocks, which have been the subject of intense research [14, 30-32]. We only need their output as they are entrained by the environment [33]. In this regard, clocks do not rely on precisely sensing the environment since only minimal coupling leads to perfect synchrony with external time. The recent environment is characterized through the $n$-dimensional vector $\mathbf{y}_{t-h,t-nh}^o = (y_{t-h}^o, y_{t-2h}^o \ldots y_{t-nh}^o)$ with the most recent $n$ values of the optimal state before the time $t$.

Explicitly, given $\mathbf{c}_t^d$, $\mathbf{c}_t^a$, and $\mathbf{y}_{t-h,t-nh}^o$, we consider $f_t = f_{DNN}^{nda}(\mathbf{y}_{t-h,t-nh}, \mathbf{c}_t^d, \mathbf{c}_t^a)$, $f_t = f_{DNN}^{nd}(\mathbf{y}_{t-h,t-nh}, \mathbf{c}_t^d)$, and $f_t = f_{DNN}^{n}(\mathbf{y}_{t-h,t-nh})$ for different values of $n$. Here, $f_{DNN}^{n}$, $f_{DNN}^{nd}$, and $f_{DNN}^{nda}$ are the outputs of the DNNs for systems without, with circadian, and with both clocks, respectively. We analyze explicitly to what extent diverse types of systems can perfectly adapt to the changing optimal state depending on the information available and the network architecture implemented. We trained the networks with the data from the years 2005 to 2014. The validation was performed with data from the years 2015 to 2020 to test the predictive capabilities with unseen data.

For systems that can only act on information about the most recent past value of the optimal state, $n = 1$, the presence of clocks significantly increases the ability to track the optimal state (Fig.

6A). The presence of clocks brings the tracking capabilities of the DNNs along the lines of the combined extrapolated and recurrent approaches (Fig. 5C). In the absence of clocks, in contrast, the capabilities of a DNN do not substantially improve the results of naïve adaptation to the last optimal value available (Fig. 4A). For systems that can act on information about the two most recent past values of the optimal state, $n = 2$, the differences between approaches are not as marked (Fig. 6B). In all the cases, the DNN outperforms the results from the linear estimates of the actionable target (Figs. 4 and 5). For large numbers of actionable past values, as for instance $n = 30$, the results are essentially the same for all the DNNs, with an RMSE of ~0.040 (Fig. 6C).

An exhaustive analysis for values up to $n = 50$ shows that all the approaches become essentially equally accurate at $n \simeq 25$ as the number of actionable values increases (Fig. 7). The most salient result is the ability of systems with daily and annual clocks to reach nearly maximum tracking accuracy, as exemplified in Fig. 6B, with just acting on to past values of the optimal state at $n = 2$. In this case, information about the recurrent environment is encoded as a function of the time of the day and of day of the year through the training of the network along the environment history.

## Discussion

Continuously adapting to the optimal state in a changing environment is essential to the survival of virtually any organism. Multiple biomolecular processes sense the environment and respond to the current condition to adapt to it. These processes include mechanisms as diverse as transcription, translation, phosphorylation, methylation, and changing multiple intracellular molecular concentrations [34]. The classical example in gene regulation, known as the *lac* operon, is the production of the enzymes needed to metabolize lactose only in the presence of lactose with glucose absent [35]. This reactive approach implies a delay in reaching the optimal state, determined by the speed of the cellular process, when bacteria switch from glucose to lactose metabolism, as shown by cells stopping growing during this transition [36]. This delay has sensing and responding components [37]. Sensing involves the transport of extracellular lactose inside the cell, its transformation into allolactose, and inhibition of the lac repressor by its binding to allolactose. Responding involves transcription and translation that leads to the production of the enzymes needed to metabolize lactose and the use of lactose as a metabolite [38]. Typically, resuming growth takes about two hours in the bacterium E. coli.

Besides reactive approaches, there are also proactive approaches, such as the regulatory, metabolic, and physiological oscillations that match daily and annual changes [13]. Therefore, multiple systems do not present a delayed response to recurrent environmental changes. In many cases, these oscillations persist even when the organisms are artificially kept under constant environments. This persistence has been attributed to the need for organisms to anticipate changes in the natural environment [33].

Here, we have shown that it is possible to continuously track the optimal state as it changes by adapting towards a combination of the current optimal state and its rate of change, which we have termed the actionable target. This quantity, and hence accurate adaptation to a changing environment, does not depend on the future values of the environment. The key limitation for accurate temporal tracking is obtaining an accurate actionable target. In general, there are inherent delays that would prevent obtaining current values and intrinsic fluctuations that would prevent obtaining precise values [39]. Therefore, the limitations for continuous adaptation are not as much as anticipating the future but obtaining reliable estimates of the present.

A key quantity besides the optimal state of the system is its rate of change. Remarkably, sensing changes, in addition to absolute values, is widely present across organisms, even in cases as simple as bacteria. The most studied example is perhaps bacterial chemotaxis, which relies on sensing temporal changes in nutrient concentrations as the bacterium moves through an inhomogeneous nutrient distribution [40]. In higher organisms, there are multiple examples of complex pathways that can sense concentrations and their changes, such as the TGF-β/BMP pathways [17]. These complex pathways can even perform complex computations [41, 42]. The computations of this sensing, however, are relayed with a delay. Therefore, the system needs to anticipate the present from delayed information.

In this context, our results show that biological clocks are not needed to anticipate future changes but to provide reliable estimates of the expected current changes that mimic historical changes. As the expected rate of change is encoded through the phase of the clock, it does not depend on sensing the environment. Our results show that, in a variable uncertain environment, biological clocks can be combined with sensing recent past values of the optimal state to increase further the reliability of the estimation of the actionable target. In the case of tracking the solar radiation on the Earth's surface, we have shown that only two recent values are needed to reach the limits of precise tracking when circadian and circannual clocks are present. The fact that the system only needs sensing two values is equivalent to sensing the environment, $y^o_{t-h}$, and its rate of change, $\frac{1}{h}(y^o_{t-h} - y^o_{t-2h})$, which are widely present capabilities of organisms. In the absence of clocks, the limit is reached at about 25 values, which is unclear how a bimolecular system would record, act upon, and process such an amount of temporal information [43]. In contrast, actionable forecasting can be implemented efficiently through biological functions that rely on oscillatory behavior.

## Methods

### Field data

We retrieved the data from the PVGIS v5.2 database [22] using the web API command 'https://re.jrc.ec.europa.eu/api/v5_2/seriescalc?lat=45&lon=0&raddatabase=PVGIS-ERA5'. Data from the years 2015 to 2020 was used for validation of the approach with numerical estimates of the actionable target and for performance assessment of DINN implementation. The data from the years 2005 to 2014 was used for training the DNN weights.

### Convolutional layer kernel parametrization

We considered the integral representation of the dynamics through Eq. (2) with a time-independent adaptation rate constant and without the transient, which after a change of variable in the convolution leads to $y_t = \int_0^t e^{-b_0(t-s)} b_0 f_s ds = \int_0^t e^{-b_0 s} b_0 f_{t-s} ds$. Using the trapezoidal rule, the integral is approximated as $y_t \simeq b_0 \sum_{i=0}^M \frac{1}{2}\left(e^{-b_0 ih} f_{t-ih} + e^{-b_0(i+1)h} f_{t-(i-1)h}\right) \simeq \sum_{i=0}^M w_i f_{t-ih}$, which defines the kernel $w_i$. We selected $b_0 = \frac{1}{3h}$ as a typical value, and $M = 25$ so that $\frac{1}{2} e^{-b_0(M+1)h} = 8.6 \times 10^{-5}$ and subsequent terms for longer delays are negligible.

### Computational Implementation

The overall approach was implemented in Keras [44] with TensorFlow [45]. For the DINN, we considered a DNN with 4 dense layers, each containing 16 neurons with the Exponential Linear Unit (ELU) function activation. As input of the DNN, we used values of the vectors $\mathbf{y}^o_{t-h,t-nh}$, $\mathbf{c}^d_t$, and $\mathbf{c}^a_t$. The output of the DNN was combined into a single-neuron linear layer, which provided the actionable target $f_t$ used as the input of the convolutional layer that implements the dynamics. The MSE between the output of the convolutional layer, $y_t$, and the optimal state, defined as the normalized solar irradiance, $y^o_t$, was used as a loss function. Training of the overall network was performed over 365 intervals 744-hour long as a single batch. The intervals were chosen randomly from the years 2005 to 2014. Validation was performed over 219 intervals 744-hour long chosen randomly from the years 2015 to 2020. For the numerical estimation of the actionable target, we considered the estimated value of $f_t$ as the input of the convolutional layer. No training was involved. Validation was performed exactly as for the DINN.

# References

1. Hartwell, L.H., Hopfield, J.J., Leibler, S., and Murray, A.W. (1999). From molecular to modular cell biology. Nature *402*, C47-52.

2. Yamada, T., and Bork, P. (2009). Evolution of biomolecular networks: lessons from metabolic and protein interactions. Nat Rev Mol Cell Biol *10*, 791-803.

3. Balazsi, G., van Oudenaarden, A., and Collins, J.J. (2011). Cellular decision making and biological noise: from microbes to mammals. Cell *144*, 910-925.

4. May, R.M. (2006). Network structure and the biology of populations. Trends in Ecology & Evolution *21*, 394-399.

5. Tkačik, G., Callan, C.G., and Bialek, W. (2008). Information flow and optimization in transcriptional regulation. Proceedings of the National Academy of Sciences *105*, 12265-12270.

6. Kussell, E., and Leibler, S. (2005). Phenotypic Diversity, Population Growth, and Information in Fluctuating Environments. Science *309*, 2075 - 2078.

7. Wu, X., Zhu, X., Wu, G.-Q., and Ding, W. (2014). Data mining with big data. IEEE transactions on knowledge data engineering *26*, 97-107.

8. Schwenzer, M., Ay, M., Bergs, T., and Abel, D. (2021). Review on model predictive control: An engineering perspective. The International Journal of Advanced Manufacturing Technology *117*, 1327-1349.

9. Chait, R., Ruess, J., Bergmiller, T., Tkačik, G., and Guet, C.C. (2017). Shaping bacterial population behavior through computer-interfaced control of individual cells. Nature Communications *8*, 1535.

10. Vilar, J.M.G. (2019). Winning the Big Data Technologies Horizon Prize: Fast and reliable forecasting of electricity grid traffic by identification of recurrent fluctuations. arXiv preprint arXiv:1902.04337.

11. Lugagne, J.-B., Blassick, C.M., and Dunlop, M.J. (2024). Deep model predictive control of gene expression in thousands of single cells. Nature Communications *15*, 2148.

12. Parker, R.S., Doyle, F.J., and Peppas, N.A. (1999). A model-based algorithm for blood glucose control in type I diabetic patients. IEEE Transactions on biomedical engineering *46*, 148-157.

13. Winfree, A.T. (2001). The Geometry of Biological Time, (Springer, New York, NY).

14. Patke, A., Young, M.W., and Axelrod, S. (2020). Molecular mechanisms and physiological importance of circadian rhythms. Nature Reviews Molecular Cell Biology *21*, 67-84.

15. Ouyang, Y., Andersson, C.R., Kondo, T., Golden, S.S., and Johnson, C.H. (1998). Resonating circadian clocks enhance fitness in cyanobacteria. Proceedings of the National Academy of Sciences *95*, 8660-8664.

16. Lambert, G., and Kussell, E. (2014). Memory and Fitness Optimization of Bacteria under Fluctuating Environments. PLoS Genetics *10*, e1004556.

17. Vilar, J.M.G., Jansen, R., and Sander, C. (2006). Signal processing in the TGF-beta superfamily ligand-receptor network. PLoS Comput Biol *2*, e3.

18. Salvy, P., and Hatzimanikatis, V. (2021). Emergence of diauxie as an optimal growth strategy under resource allocation constraints in cellular metabolism. Proceedings of the National Academy of Sciences *118*, e2013836118.

19. Lan, G., Sartori, P., Neumann, S., Sourjik, V., and Tu, Y. (2012). The energy–speed–accuracy trade-off in sensory adaptation. Nature Physics *8*, 422-428.

20. McClung, C.R. (2006). Plant Circadian Rhythms. The Plant Cell *18*, 792-803.

21. Vilar, J.M.G., and Saiz, L. (2023). Dynamics-informed deconvolutional neural networks for super-resolution identification of regime changes in epidemiological time series. Sci Adv *9*, eadf0673.

22. Huld, T., Müller, R., and Gambardella, A. (2012). A new solar radiation database for estimating PV performance in Europe and Africa. Solar Energy *86*, 1803-1815.

23. Winters, P.R. (1960). Forecasting sales by exponentially weighted moving averages. Management science *6*, 324-342.

24. Okumura, S., Gines, G., Lobato-Dauzier, N., Baccouche, A., Deteix, R., Fujii, T., Rondelez, Y., and Genot, A.J. (2022). Nonlinear decision-making with enzymatic neural networks. Nature *610*, 496-501.

25. Kim, J., Hopfield, J., and Winfree, E. (2004). Neural network computation by in vitro transcriptional circuits. Advances in neural information processing systems *17*.

26. Cherry, K.M., and Qian, L. (2018). Scaling up molecular pattern recognition with DNA-based winner-take-all neural networks. Nature *559*, 370-376.

27. Evans, C.G., O'Brien, J., Winfree, E., and Murugan, A. (2024). Pattern recognition in the nucleation kinetics of non-equilibrium self-assembly. Nature *625*, 500-507.

28. Lincoln, G., and Hazlerigg, D. (2019). Mammalian circannual pacemakers. Society of Reproduction and Fertility supplement *67*, 171-186.

29. Visser, M.E., Caro, S.P., Van Oers, K., Schaper, S.V., and Helm, B. (2010). Phenology, seasonal timing and circannual rhythms: towards a unified framework. Philosophical Transactions of the Royal Society B: Biological Sciences *365*, 3113-3127.

30. Vilar, J.M.G., Kueh, H.Y., Barkai, N., and Leibler, S. (2002). Mechanisms of noise-resistance in genetic oscillators. Proc Natl Acad Sci U S A *99*, 5988-5992.

31. Pittayakanchit, W., Lu, Z., Chew, J., Rust, M.J., and Murugan, A. (2018). Biophysical clocks face a trade-off between internal and external noise resistance. Elife *7*, e37624.

32. Meyer, P., Saez, L., and Young, M.W. (2006). PER-TIM Interactions in Living Drosophila Cells: An Interval Timer for the Circadian Clock. Science *311*, 226-229.

33. Golombek, D.A., and Rosenstein, R.E. (2010). Physiology of Circadian Entrainment. Physiological Reviews *90*, 1063-1102.

34. Kolch, W., Halasz, M., Granovskaya, M., and Kholodenko, B.N. (2015). The dynamic control of signal transduction networks in cancer cells. Nature Reviews Cancer *15*, 515-527.

35. Jacob, F., and Monod, J. (1961). Genetic regulatory mechanisms in the synthesis of proteins. Journal of Molecular Biology *3*, 318-356.

36. Monod, J. (1949). The growth of bacterial cultures. Annual review of microbiology *3*, 371-394.

37. Vilar, J.M.G., Guet, C.C., and Leibler, S. (2003). Modeling network dynamics: the lac operon, a case study. J Cell Biol *161*, 471-476.

38. Vilar, J.M.G., and Saiz, L. (2013). Systems biophysics of gene expression. Biophys J *104*, 2574-2585.

39. Saiz, L., and Vilar, J.M. (2006). Stochastic dynamics of macromolecular-assembly networks. Mol Syst Biol *2*, 2006 0024.

40. Barkai, N., and Leibler, S. (1997). Robustness in simple biochemical networks. Nature *387*, 913-917.

41. Antebi, Y.E., Linton, J.M., Klumpe, H., Bintu, B., Gong, M., Su, C., McCardell, R., and Elowitz, M.B. (2017). Combinatorial Signal Perception in the BMP Pathway. Cell *170*, 1184-1196 e1124.

42. Vilar, J.M.G., and Saiz, L. (2017). Computing at the Front-End by Receptor Networks. Cell Syst *5*, 316-318.

43. Tjalma, A.J., Galstyan, V., Goedhart, J., Slim, L., Becker, N.B., and ten Wolde, P.R. (2023). Trade-offs between cost and information in cellular prediction. Proceedings of the National Academy of Sciences *120*, e2303078120.

44. Chollet, F. (2018). Deep learning with Python, (Shelter Island, NY: Manning Publications Co.).

45. Abadi, M., Barham, P., Chen, J., Chen, Z., Davis, A., Dean, J., Devin, M., Ghemawat, S., Irving, G., and Isard, M. (2016). TensorFlow: a system for Large-Scale machine learning. 12th USENIX symposium on operating systems design and implementation (OSDI 16), 265-283.

## Acknowledgments

J.M.G.V. acknowledges support from Ministerio de Ciencia e Innovación under grant PID2021-128850NB-I00 (MCI/AEI/FEDER, UE).

## Author contributions

J.M.G.V and L.S. conceived, designed, and performed the research.

# Figures

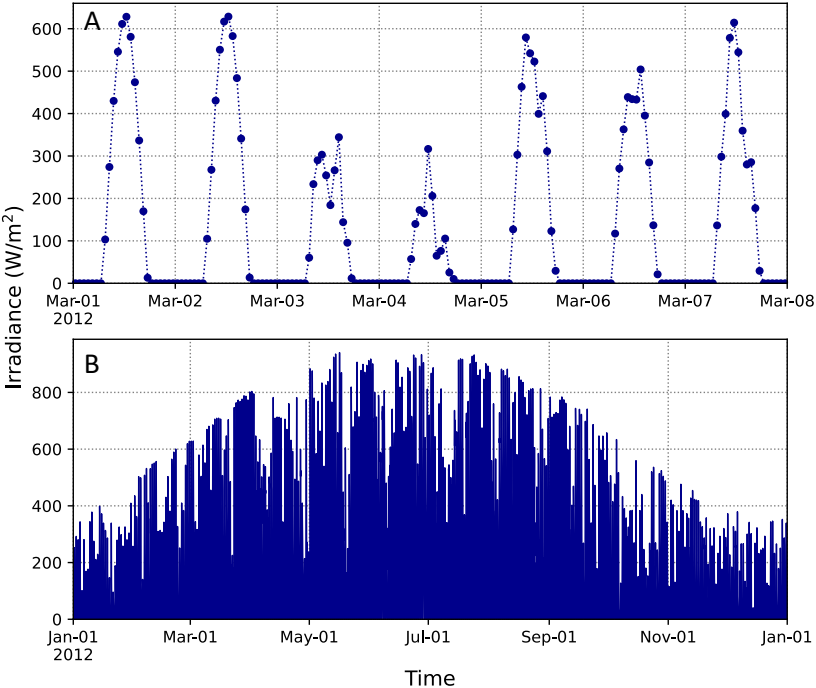

**Figure 1. Solar radiation on the Earth's surface at latitude 45° N and longitude 0° E illustrates the recurrent daily and annual fluctuations.** The values from the PVGIS v5.2 database [22] are shown for a period of a week to illustrate daily fluctuations (A) and a year to illustrate seasonal variability (B).

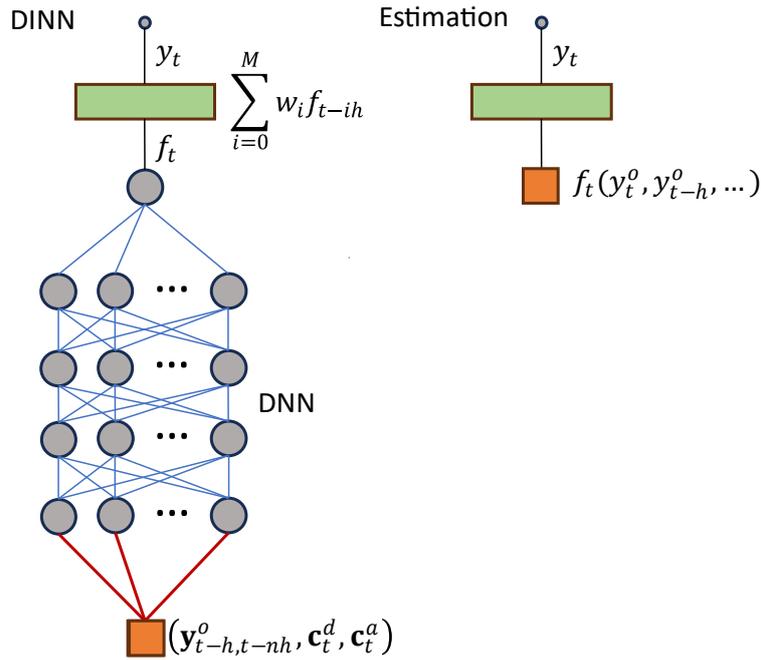

**Figure 2. Implementation of continuous adaptation as a dynamics-informed neural network (DINN).** The panel on left shows, from bottom to top, the structure of the DINN. The input, depicted as an orange square, consists of values of the vectors $\mathbf{y}^o_{t-h,t-nh}$, $\mathbf{c}^d_t$, and $\mathbf{c}^a_t$, which are connected to the lower layer of the DNN (red lines). The DNN consists of 4 dense layers, each containing 16 neurons with the Exponential Linear Unit (ELU) function activation. The output of the DNN was combined into a single-neuron linear layer, which provided the actionable target $f_t$. Blue lines connecting dense neurons (grey circles) represent trainable weights. The actionable target is used as the input of the convolutional layer (green rectangle) that implements the dynamics. The output of the convolutional layer is the state of the system $y_t$. The panel on the left illustrates the approach with the numerical estimation of the actionable target. We considered the estimated value of $f_t$ (orange square) as the input of the convolutional layer.

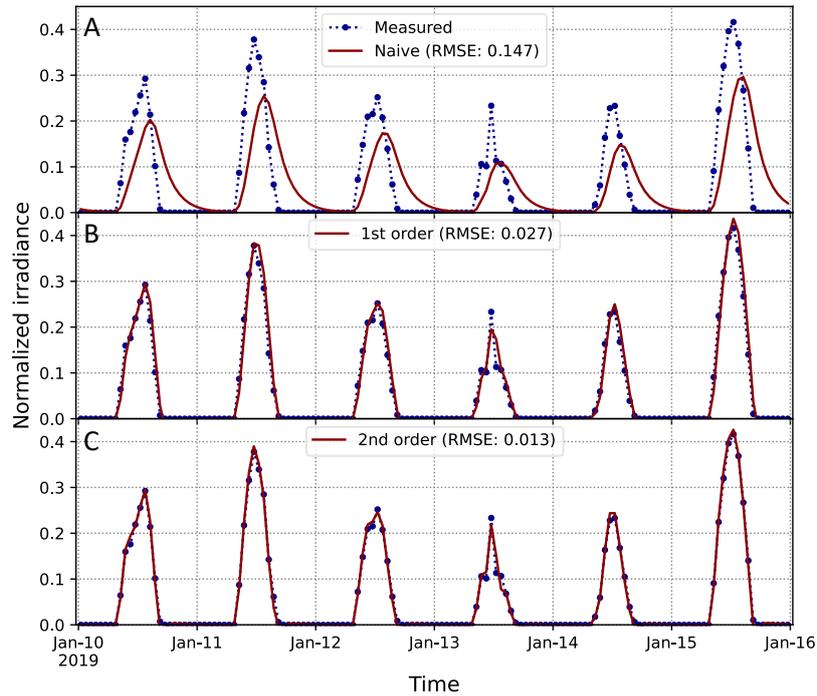

**Figure 3. Adaptation towards the actionable target leads to precise tracking even with discretely sampled data.** The dotted blue line shows normalized measured values from the PVGIS v5.2 database [22]. Solid red lines show the results for the adaptation towards the current optimal state (A), the current actionable target with a first-order discrete backward approximation of the derivative (B), and the current actionable target with a second-order discrete backward approximation of the derivative (C). The RMSE for the years 2015-2020 between the measured (dotted blue line) and tracked values (solid lines) are shown in the legends in each panel.

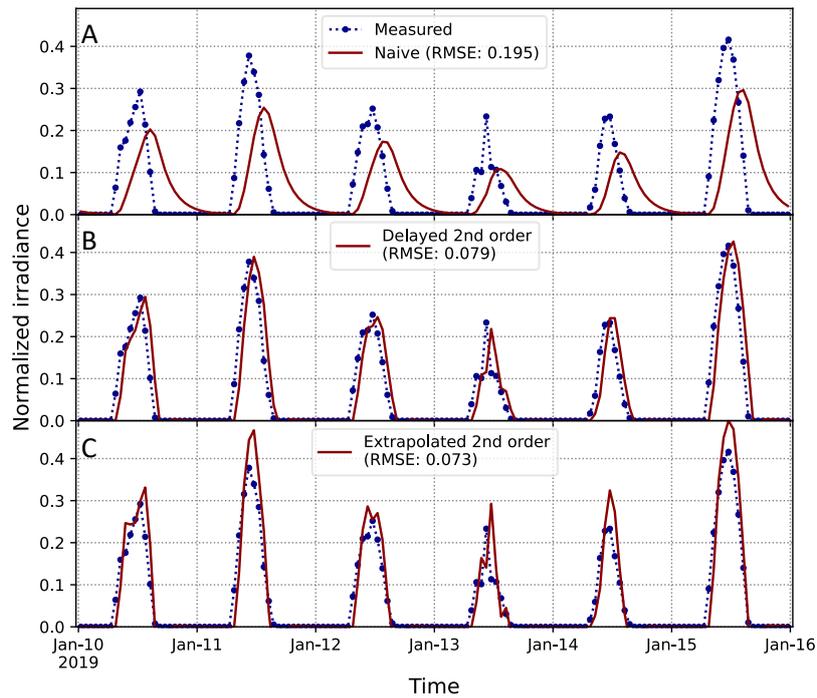

**Figure 4. Delayed relay of the actionable target prevents perfect tracking but improves naïve adaptation.** The dotted blue line shows normalized measured values from the PVGIS v5.2 database [22]. Solid red lines show the results for the adaptation towards the 1-hour-delayed optimal state (A), the 1-hour-delayed actionable target (B), and the extrapolation to current time from the 1-hour-delayed actionable target (C). In both cases, the actionable target was computed with a second-order discrete backward approximation of the derivative. The RMSE for the years 2015-2020 between the measured (dotted blue line) and tracked values (solid lines) are shown in the legends in each panel.

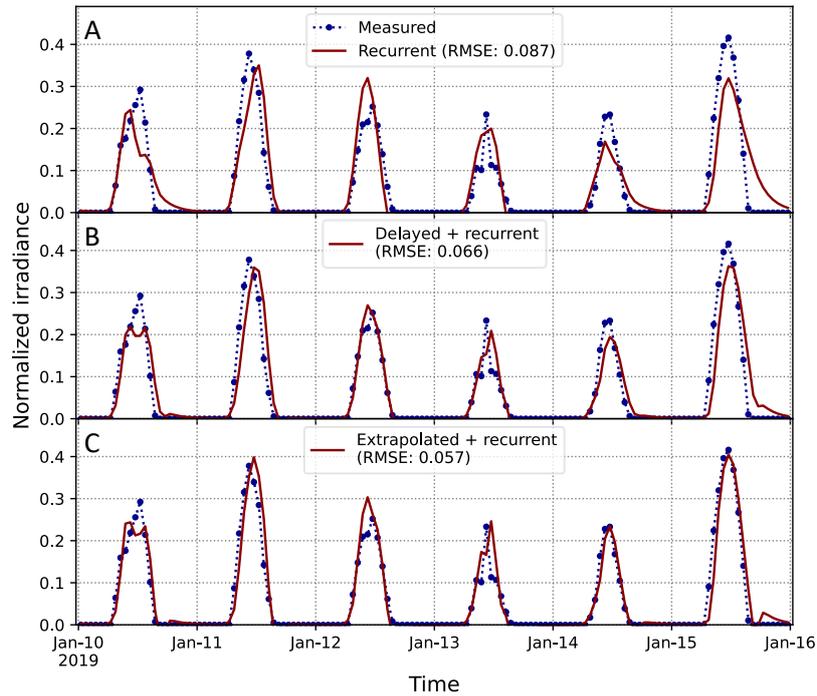

**Figure 5. Estimation of the actionable target with long-term recurrent changes improves tracking.** The dotted blue line shows normalized measured values from the PVGIS v5.2 database [22]. Solid red lines show the results for the adaptation towards the current time estimate of the actionable target with values of the previous day (A), the average of panel A and Fig. 4B actionable targets (B), and the average of panel A and Fig. 4C actionable targets (C). The RMSE for the years 2015-2020 between the measured (dotted blue line) and tracked values (solid lines) are shown in the legends in each panel.

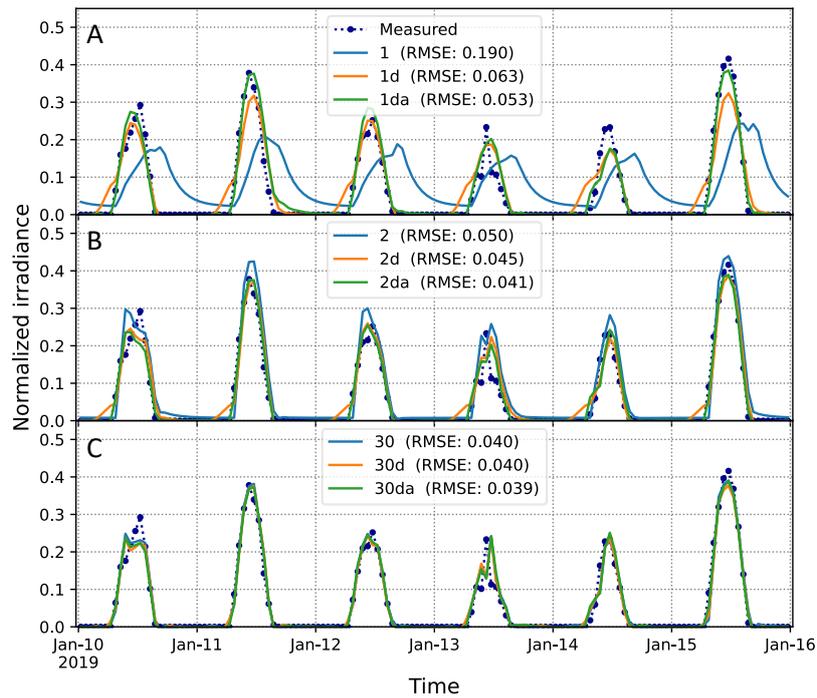

**Figure 6. A dynamics-informed neural network with daily and yearly clocks reaches the limit of predictability with just two recent past values of the optimal state.** The dotted blue line shows normalized measured values from the PVGIS v5.2 database [23]. Solid lines show the results for the adaptation towards the estimate of the actionable target using a dynamics-informed neural network (DINN) for systems without (blue), with daily (orange), and with daily and annual (green) clocks for 1 (A), 2 (B), and 30 (C) actionable past values of the optimal state. The RMSE for the years 2015-2020 between the measured (dotted blue line) and tracked values (solid lines) are shown in the legends in each panel.

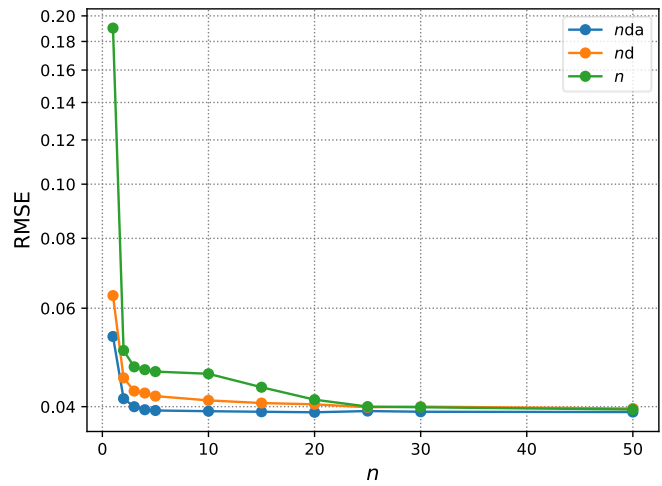

**Figure 7. Dynamics-informed neural network approaches without, with daily, and with daily and annual clocks become essentially equally accurate at $n \simeq 25$ as the number of actionable past values of the environment increases.** The RMSE for the years 2015-2020 between the measured and tracked values is shown as a function of the number of past actionable values, denoted by $n$, for systems without (blue), with daily (orange), and with daily and annual (green) clocks.